\documentstyle[twoside,fleqn,psfig,espcrc2]{article}%

\newcommand{\be}{\begin{equation}}
\newcommand{\ee}{\end{equation}}
\newcommand{\bea}{\begin{eqnarray}}
\newcommand{\eea}{\end{eqnarray}}

\def\opket#1#2 {  {#1} |{#2}\rangle }

\def\dydxh#1#2{ \partial{#1} / \partial {#2} }

\def\det{{\rm det}\,}

\def\Im{{\rm Im}}

\def\K2{{\cal K}}

\def\mat#1#2#3#4{  \left( \matrix{ {#1} & {#2} \cr
                                   \noalign{\vskip3pt}
                                   {#3} & {#4} \cr    } \right) }

\def\Re{{\rm Re}}

\def\Tr{{\rm Tr}}

\begin{document}                                                        
\renewcommand{\refname}{\normalsize\bf References}
\title{Regular Tunnelling Sequences in Mixed Systems}

\author{%
	Stephen C. Creagh%
	\address{Division of Theoretical Mechanics,
                 School of Mathematical Sciences\\
                 University of Nottingham, NG7 2RD, UK}
        \,and
        Niall D. Whelan%
        \address{Department of Physics and Astronomy,
	         McMaster University,
		 Hamilton, Ontario, Canada L8P 2E7}
}
\begin{abstract}
\hrule
\mbox{}\\[-0.2cm]

\noindent{\bf Abstract}\\

We show that the pattern of tunnelling rates can display a vivid and
regular pattern when the classical dynamics is of mixed
chaotic/regular type. We consider the situation in which the dominant
tunnelling route connects to a stable periodic orbit and this orbit is
surrounded by a regular island which supports a number of quantum
states. We derive an explicit semiclassical expression for the
positions and tunnelling rates of these states by use of a
complexified trace formula.

\hrule
\end{abstract}

\maketitle

Tunnelling in systems whose classical limit displays a mixture of
chaotic and integrable behaviour \cite{cat1,cat2,cat3,cat4} is often
quite complex and impossible to predict analytically. Much attention
has been paid recently, for example, to the regime of chaos-assisted
tunnelling \cite{cat2} in which dynamical tunnelling occurs between
quasimodes supported in integrable island-chains embedded in a chaotic
sea. By contrast, we report on a remarkably ordered structure that appears
in the tunnelling behaviour of a particular kind of mixed system and
give analytical estimates for the corresponding tunnelling rates. It
is distinct from the case of chaos-assisted tunnelling because
tunnelling is through an energetic barrier rather than through
dynamical barriers such as KAM tori. The special feature of these
systems is that the complex orbit which defines the optimal tunnelling
route across the barrier connects to a stable periodic orbit at the
centre of an island chain (which is generally embedded in a chaotic
sea).

\begin{figure}[h]
\vspace*{0.4in}
\hspace*{0.0in}\psfig{figure=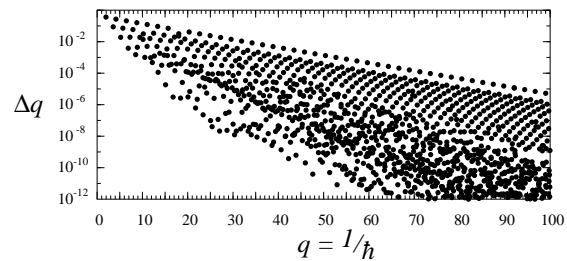,height=1.1in}
\vspace*{-0.6in}
\caption
{\small Each dot represents a splitting doublet. The horizontal
coordinate is the mean value of $q=1/\hbar$ and the vertical
coordinate is the splitting.}
\label{interesting}
\vspace*{-0.2in}
\end{figure}

The ordered nature of these tunnelling rates is immediately evident in
Fig.~\ref{interesting} where we show the numerically obtained
splittings between quasi-doublets of the double-well potential $V(x,y)
= (x^2-1)^4 + x^2y^2 + 2y^2/5$. We have held the energy fixed (at
$E=9/10$) and found the values of $q=1/\hbar$ for which this is an
energy level. The resulting spectrum of $q$ values is equivalent in
most respects to a standard energy-spectrum, with the advantage that
the classical dynamics is fixed throughout.

The largest splittings in Fig.~\ref{interesting} are highly ordered,
forming a regular progression of families which grow larger in number
higher in the spectrum. These correspond to states supported near the
centre of the island and we will offer a simple analytical prediction
for them. The smallest splittings in Fig.~\ref{interesting} form a
disordered jumble. These correspond to states supported in the chaotic
sea and dynamically excluded from the main tunnelling route.

To analyse the ordered sequence we use a method developed in
\cite{us1,us2} and used until now primarily to understand
predominantly chaotic systems. We first present the analysis for the
case that $q=1/\hbar$ is held fixed and an energy-spectrum is
computed. We then give the simple extension for the fixed energy
$q$-spectrum, appropriate for the results of
Fig.~\ref{interesting}. We call the mean energy of the $n$'th
doublet $E_n$ and the corresponding splitting $\Delta E_n$ and define
the following dimensionless spectral function
\be \label{fdef}
f(E,q) = \sum_n \Delta E_n \delta(E-E_n).
\ee
There is an analogous definition for metastable wells which have
extremely narrow resonances, in which the widths $\Gamma_n$ play the
role of splittings. Such a system was studied in \cite{delos} for a
fully integrable system; the structure of the spectrum was like that
shown here except with no irregular jumble at the bottom. While
similar in outline, the detailed method of analysis was rather
different, making use of the action-angle variables which exist in
that situation.

In \cite{us1,us2} we approximate (\ref{fdef}) semiclassically as a sum
over complex tunnelling orbits which traverse the barrier (in analogy
to Gutzwiller's formula for the density of states using real orbits
\cite{gutz}). We shall consider the special case in which there is an
additional reflection symmetry such that the dominant tunnelling route
lies on the symmetry axis so that it connects smoothly to a real
periodic orbit. We then identify three distinct contributions to
$f(E,q)$. There is the so-called instanton which has a purely
imaginary action $iK_0$, lives under the barrier and runs along the
symmetry axis between the classical turning points. It is important
for determining the mean behaviour of the splittings but does not
affect the fluctuation effects which we are trying to capture
here. The second contribution comes from orbits which execute real
dynamics along the real periodic orbit lying on the symmetry axis in
addition to the instanton dynamics beneath the well. We imagine an
orbit which starts at one of the turning points, executes $r$
repetitions of the real periodic orbit and then tunnels along the
instanton path to finish at the other turning point. This orbit has a
complex action $S=rS_0 + iK_0$ where $S_0$ is the real action of the
real periodic orbit. The contribution to $f(E,q)$ is given by
\cite{us1}
\begin{equation} \label{fosc}
f_{\rm osc}(E,q) = {2 \over \pi} \Re \sum_{r=1}^\infty
{e^{-qK_0+riqS_0} \over \sqrt{-{\rm det}(W_0M_0^r-I)}}.
\end{equation}
The matrices $W_0$ and $M_0$ are the monodromy matrices of the
instanton and of the real periodic orbit respectively; the composite
orbit has a monodromy matrix which is simply a product of these. The
third contribution, discussed in \cite{us2}, comes from homoclinic
orbits which explore the real wells far away from the symmetry axis.
They play no role in the present discussion.

For fully developed chaos, all periodic orbits are unstable. The
denominator of (\ref{fosc}) then decays exponentially with $r$ and
large repetitions are numerically unimportant. If the orbit is stable,
however, there is no corresponding suppression. Singularities
corresponding to distinct states arise when the expression is
summed. This follows very closely the analogous development of Miller
and Voros \cite{voros,miller,Richens} for the Gutzwiller trace formula
when there is a stable orbit, except here we find splittings in
addition to the positions of energies. (As was done in one dimension
by Miller \cite{miller2}.)

In the stable case, $M_0$ has eigenvalues $e^{\pm i\alpha}$ on the
unit circle.  (In higher dimensions, there would be a number of such
eigenvalues and the theory would be generalised accordingly.) Let the
diagonal matrix elements of $W_0$ in the complex eigenbasis of $M_0$
be $A$ and $B$ so that
\begin{eqnarray}\label{defAB}
-{\rm det}(W_0M_0^r-I)&=&\Tr \, W_0M_0^r-2 \nonumber\\
&=&   Ae^{ir\alpha} + B e^{-ir\alpha}-2.
\end{eqnarray}
Since the instanton's period is imaginary, complex conjugation acts as
a time-reversal operation and we find that $W_0^*=W_0^{-1}$. We
therefore conclude that 
$(\Tr W_0 M_0^r)^*=\Tr(W_0M_0^{-r})^{-1} = \Tr W_0M_0^{-r}$ (the
latter equality holds because every symplectic matrix is conjugate to
its inverse). Comparing this with (\ref{defAB}) we conclude that $A$
and $B$ are real; we discuss how they are computed and offer a
geometrical interpretation in the appendix.

We now make use of the generating function of the Legendre polynomials
to conclude
\be
{1\over \sqrt{-{\rm det}(WM_0^r-I)}} 
	=
	\sum_{k=0}^\infty e^{i(k+1/2)r\alpha}
 		\frac{Q_k(AB)}{B^{k+1/2}}
\ee
where we assume without loss of generality that $B$ 
is the larger in magnitude of $(A,B)$ and we let $Q_k (z)$ denote the
polynomial 
\be
	Q_k (z) = z^{k/2} P_k(z^{-1/2}).
\ee
Using this in (\ref{fosc}) and summing the resulting geometric
series in $r$ for each $k$ we get,
\be \label{bigresult}
f_{\rm osc}(E) = {2e^{-qK_0} \over \pi} \Re \sum_{k=0}^\infty
\frac{a_k}{e^{-i\Phi_k}-1},
\ee
where we have defined
\begin{eqnarray}
a_k& = & \frac{Q_k(AB)}{B^{k+1/2}}\nonumber\\
\Phi_k & = & qS_0 + (k+1/2)\alpha.
\end{eqnarray}
Semiclassical energy levels are found when the distribution above has
poles and are implicit solutions $E_{mk}$ of $\Phi_k=2\pi m$. From the
residues we recover estimates of the corresponding splittings.  This
is a form of torus quantisation in which $k$ is a transverse quantum
number, treated in harmonic approximation, and $m$ counts nodes along
the orbit. The corresponding states are localised on the tori
surrounding the stable periodic orbit and we find that their
respective tunnelling rates are much larger than those of other
states.

Note that near $E_{mk}$ we can write
\begin{equation}
{1\over e^{-i\Phi_k}-1} \approx {i\over qT_0} {1\over E-E_{mk}},
\end{equation}
where we have used that the period is $T_0=\dydxh{S_0}{E}$. Using the
standard identity
\begin{equation}
\Im{1\over E-E_{mk}} = -\pi\delta(E-E_{mk})
\end{equation}
we conclude
\be
\Delta E_{mk} = {2\hbar \over T_0} e^{-K_0/\hbar} a_k(W_0,M_0),
\ee
where the notation now stresses that $a_k$ depends on the transverse
dynamics of the instanton and its real extension through the monodromy
matrices $(W_0,M_0)$.  All classical quantities are evaluated at
energy $E_{mk}$. This formula has the same form as in one-dimensional
tunnelling \cite{miller2} except for the additional factor
$a_k(W_0,M_0)$. We will find this factor to be of order unity when $k$
is small and to decrease as $k$ increases.

In \cite{us1} we reported some specific numerical results for the
energy quantisation. We now extend this result to the the
$q$-spectrum, the advantage being that the classical quantities $K_0$,
$W_0$ and $M_0$ are constant. The function $f(E,q)$ can equally well
be interpreted as a function of $q$ at fixed $E$ so that
(\ref{bigresult}) still applies. Solving for the poles and residues as
above, we conclude
\bea\label{qresult}
q_{mk}        & = & \left(2m\pi - (k+1/2)\alpha\right)/S_0 \nonumber\\
\Delta q_{mk} & = & {2\over S_0} e^{-q_{mk}K_0} a_k(W_0,M_0).
\eea
The sequences of Fig.~\ref{interesting} can now be interpreted in
terms of the quantum numbers $m$ and $k$.

The central states with $k=0$ correspond to the largest splittings,
about $35$ times larger than the local average. In
Fig.~\ref{interesting} they are the uppermost curve of points (along
which $m$ varies). Keeping $m$ fixed and letting $k$ increase one gets
a sequence, which appears as a left-sloping shoulder in
Fig.~\ref{interesting}, along which $q$ and $\Delta q$ decrease.

\begin{figure}[h]
\vspace*{0.4in}
\hspace*{0.0in}\psfig{figure=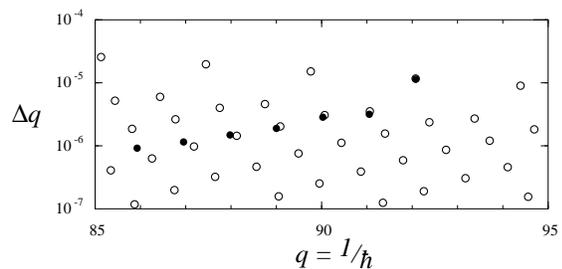,height=1.1in}
\vspace*{-0.6in}
\caption
{\small The open circles are numerically-computed splittings. The
filled circles are the semiclassical predictions for the family with
$m=40$ and $0\leq k\leq6$.
}
\label{itworks}
\vspace*{-.4cm}
\end{figure}

In Fig.\ref{itworks} we show a subset of the spectrum with the
semiclassical predictions for $m=40$ and a sequence of $k$ values
(using $A=-0.861$, $B=4.043$ and $\alpha=2.783$.)  Clearly the small
$k$ states are well reproduced. Our analysis essentially extrapolates
the properties of the central periodic orbit to the entire island
\cite{voros}, and this is less accurate for the large $k$ states
which are localised further from the periodic orbit. Reproducing the
large $k$ values would require a more sophisticated analysis
\cite{Richens}, although our formalism does at least capture the
correct qualitative behaviour of these states. The irregular jumble of
splittings at the bottom of the figure corresponds to states in the
chaotic part of phase space. No simple theory exists for them though
one could well imagine that the formalism of chaos-assisted
tunnelling, in particular the interplay between regular and chaotic
states, might be of use in describing them.

As remarked, the number of well-defined $k$ states increases as we go
up in the spectrum. This is because the number of states which the
regular island can support increases as $\hbar$ decreases. Also, there
are occasional irregularities in the lattice of regular states, for
example the $k=0$ state near $q=18$. These are due to near
degeneracies between the regular state and some other state --- either
another regular one or a chaotic one. The actual eigenstates are then
strongly mixed and hence so are the tunnelling rates.

\vspace*{12pt}
\noindent{\bf Appendix}\\

To apply Eq.~(\ref{qresult}) in practice one needs the parameters $A$
and $B$; we give here simple basis-independent expressions for them.
In particular, we note that it is not explicitly necessary to
transform $W_0$ to the eigenbasis of $M_0$.
We calculate $A$ and $B$ as the smaller and larger
respectively of
\begin{equation}
A \mbox{ or } B = \cosh\beta \pm \gamma\sinh\beta
\end{equation}
where $\Tr W_0=2\cosh\beta$ and, 
\begin{equation}
\gamma = \frac{\Im\,\Tr\,W_0 M_0}{2\sin\alpha\sinh\beta}.
\end{equation}
This is obtained by expressing $W_0=e^{-i\beta JH}$ and $M_0=e^{\alpha
JK}$ where $J$ is the unit symplectic matrix and $H$ and $K$ are real,
positive-definite, $2\times 2$ symmetric matrices normalised so that
$\det H=\det K=1$.  Expanding
\be W_0=e^{-i\beta JH}=\cosh\beta-iJH\sinh\beta
\ee
and similarly for ${M_0}^r=e^{r\alpha JK}$, one recovers (\ref{defAB})
with $A$ and $B$ as given above.

The factor $\gamma$ has the following geometric interpretation. The
action $K\to MKM^T$ of $2\times2$ symplectic matrices $M$ on symmetric
matrices $K$ can be identified with $(2+1)$-dimensional Lorentz
transformations (since the relevant Lie algebras are isomorphic), the
invariant $\det K$ playing the role of proper time. The matrices $H$
and $K$ define unit time-like $(2+1)$-vectors $X=(x,y,t)$ and
$\Xi=(\xi,\eta,\tau)$ respectively. For example
\be
H = \mat{t+x}{y}{y}{t-x},\quad t^2-x^2-y^2=1,
\ee
and similarly for $K$. One then observes that $\gamma=-\Tr JHJK/2 =
\langle X,\Xi\rangle = t\tau-x\xi-y\eta$. This can be interpreted as
the dilation factor to boost the rest-frame of $X$ to that of $\Xi$.
In particular, $\gamma>1$.

\end{document}